\documentclass{article}
\usepackage{graphicx}
\usepackage{amsmath}
\usepackage{amsfonts}
\usepackage{amssymb}

\begin{document}

\title{Deterministic Dynamics in the Minority Game}
\author{P. Jefferies, M.L. Hart and N.F. Johnson\\Physics Department, Clarendon Laboratory\\Oxford University, Oxford OX1 3PU, U.K.}
\maketitle
\begin{abstract}
The Minority Game (MG) behaves as a stochastically perturbed deterministic
system due to the coin-toss invoked to resolve tied strategies. Averaging over
this stochasticity yields a description of the MG's deterministic dynamics via
mapping equations for the strategy score and global information. The
strategy-score map contains both restoring-force and bias terms, whose
magnitudes depend on the game's quenched disorder. Approximate analytical
expressions are obtained and the effect of `market impact' discussed. The
global-information map represents a trajectory on a De Bruijn graph. For small
quenched disorder, an Eulerian trail represents a stable attractor. It is
shown analytically how anti-persistence arises. The response to perturbations
and different initial conditions are also discussed.
\end{abstract}

\vskip1cm

\vskip1cm

\newpage

\section{Introduction}

The Minority Game (MG) introduced by Challet and Zhang
\cite{Challet-Zhang_Original-MG} offers possibly the simplest paradigm for a
complex, dynamical system comprising many competing agents. Models based on
the Minority Game concept have a broad range of potential applications, for
example financial markets, biological systems, crowding phenomena and routing
problems \cite{EconPhys_Website}. There have been many studies of the
statistical properties of the MG
\cite{Challet-Zhang_Original-MG,EconPhys_Website,
Challet_Reduced-Space,Mike_Belgium, Us_Crowd-AntiCrowd,Savit_Original-MG,
Sherrington_TMG,Us_TMG,Challet_MG-Memory,Cavagna_RandomHist,heimel,
Mike_TMG_Markov,Challet_StationaryStates,Zheng_EfficientStats} which treat the
game as a quasi-stochastic system.

In this paper we examine the MG from a different perspective by treating it as
a primarily \emph{deterministic} system and then exploring the rich dynamics
which result. Our desire to look at microscopic dynamical properties, as
opposed to global statistics, is motivated by the fact that the physical
systems we are interested in modeling are only realized once (e.g. the
time-evolution of a financial market). Only limited insight is therefore
available from taking configuration averages in such cases. In addition it is
of great interest to examine transient effects such as the response of the
system to perturbations, and the mechanisms which determine the game's
trajectories in time. We find that we are able to provide a description of the
resulting deterministic dynamics via mapping equations, and can hence
investigate these important effects. The outline of the paper is as follows:
after briefly discussing the MG in the remainder of this section, Sec. 2
examines the MG as a functional map. Section 3 focuses on the effect of the
underlying (`quenched') disorder arising from unequal population of the
strategy-space. Section 4 discusses the dynamics of the game on a de Bruijn
graph. Section 5 provides the conclusions.

The most basic formulation of a MG comprises an odd number of agents $N$ who
at each turn of the game choose between two options `0' and `1'
\cite{Challet-Zhang_Original-MG,EconPhys_Website}. These options could be used
to represent buy/sell, choose road A/road B etc. The aim of the agents is
common: to choose the least-subscribed option, the `minority' group. At the
end of each turn of the game, the winning decision corresponds to the minority
group and is announced to all the agents. The agents have a \emph{memory} of
$m$ bits, hence they can recall the last $m$ winning decisions. The
\emph{global information} $\mu$\ available to each and every agent is
therefore a binary word $m$ bits long, hence $\mu$ belongs to the set
$\left\{  0,1,...P-1\right\} $ where $P=2^{m}$. In order to make a decision
about which option to choose, each agent is allotted $s$ \emph{strategies } at
the outset of the game, which cannot be altered during the game. Each strategy
$R$\ maps every possible value of $\mu$ to a prediction $a_{R}^{\mu}%
\in\left\{  -1,1\right\}  $ where $-1\Rightarrow\left(  \text{option
0}\right)  $ and $1\Rightarrow\left(  \text{option 1}\right)  $. There are
$2^{2^{m}}$\ different possible binary strategies. \ However, many of the
strategies in this space are similar to one another, i.e. they are separated
by a small Hamming distance. It has been shown \cite{Challet_Reduced-Space}
that the principle features of the MG are reproduced in a smaller
\emph{Reduced Strategy Space} (RSS)\ of $2^{m+1}$\ strategies, in which any
two strategies are separated by a Hamming distance of 2$^{m}$\ or $2^{m-1}$,
i.e. the two strategies are \emph{anti-correlated} or \emph{un-correlated} respectively.

The agents follow the prediction of their historically best-performing
strategy. They measure this performance by rewarding strategies with the
correct mapping of global information to winning decision, and penalizing
those with an incorrect mapping. Strategies are scored in this manner
irrespective of whether they are played. As each agent will reward and
penalize the same strategy in the same way, there is a common set of strategy
scores which are collected together to form the \emph{strategy score vector}
$\underline{S}$. The common perception of a strategy's success or failure will
lead to agents deciding to use or avoid the same strategy in groups - this
leads to crowd behavior as analyzed in Refs.
\cite{Mike_Belgium,Us_Crowd-AntiCrowd}.

\bigskip

\section{MG as a functional map}

The Minority Game is often introduced heuristically as a set of rules
determining the update of the agents' strategies and the global information.
\ It can however easily be cast into a functional map which reproduces the
game when iterated. Moreover, this functional map can be iterated
\emph{without} having to keep track of the labels for individual agents. We
achieve this by introducing a formalism which groups together agents who hold
the same combination of strategies, and hence respond in an identical way to
all values of the global information set $\mu=\{0,1...P-1\}$. This grouping is
achieved via the tensor \underline{\underline{$\Omega$}} which is initialized
at the outset of the game and quantifies the particular quenched disorder for
that game \cite{Mike_Belgium}. \underline{\underline{$\Omega$}} is
$s$-dimensional with rows/columns of length $2P$ (in the RSS) such that entry
\underline{\underline{$\Omega$}}$_{R1,R2,...}$ is the number of agents holding
strategies $\{R1,R2,...\}$. The entries of \underline{\underline{$\Omega$}}
(and also of the strategy score vector \underline{$S$}) are ordered by
increasing decimal equivalent. For example, strategies from the RSS for $m=2$
are ordered $\{0000,0011,0101,0110,...\}$, therefore strategy $R$ is
anti-correlated to strategy $2P+1-R$. \underline{\underline{$\Omega$}} is
randomly filled with uniform probability such that
\[
\sum_{R,R^{\prime},...}\underline{\underline{\Omega}}_{R,R^{\prime},...}=N
\]
It is useful to construct a configuration of this tensor, \underline
{\underline{$\Psi$}}, which is symmetric in the sense that \underline
{\underline{$\Psi$}}$_{\{R1,R2,...\}}=$ \underline{\underline{$\Psi$}%
}$_{p\{R1,R2,...\}}$ where $p\{R1,R2,...\}$ is any permutation of strategies
$R1,R2,...$ . For $s=2$ we let \underline{\underline{$\Psi$}}$=\frac{1}%
{2}\left(  \underline{\underline{\Omega}}+\underline{\underline{\Omega}%
}^{\text{T}}\right)  $ \cite{countpsi}. Now we proceed to a formula for the
attendance $A$ of the MG (i.e. the sum of all the agents' predictions and
hence actions):
\begin{equation}
A=\underline{a}^{\mu}\cdot\underline{n}=\sum_{R=1}^{2P}a_{R}^{\mu}%
n_{R}\label{Eq Attendance Basic}%
\end{equation}
where $a_{R}^{\mu}$ is the response of strategy $R$ to global information
$\mu$ and $n_{R}$ is the number of agents playing strategy $R$. We can define
$n_{R}$ in terms of the strategy score vector \underline{$S$} and
\underline{\underline{$\Psi$}} and hence rewrite Eq. \ref{Eq Attendance
Basic}\ to give the following for $s=2$:
\begin{align}
A\left[  \underline{S},\mu\right]   &  =\sum_{R=1}^{2P}a_{R}^{\mu}%
\sum_{R^{\prime}=1}^{2P}\left(  1+\operatorname{sgn}\left[  S_{R}%
-S_{R^{\prime}}\right]  \right)  \underline{\underline{\Psi}}_{R,R^{\prime}%
}\label{Eq Attendance}\\
&  +\sum_{R\neq R^{\prime}}^{2P}a_{R}^{\mu}\delta_{S_{R},S_{R^{\prime}}%
}\left(  \operatorname{bin}\left[  2\underline{\underline{\Psi}}_{R,R^{\prime
}},\frac{1}{2}\right]  -\underline{\underline{\Psi}}_{R,R^{\prime}}\right)
\nonumber
\end{align}
where $\operatorname{bin}\left[  n,p\right]  $ is a sample from a binomial
distribution of $n$ trials with probability of success $p$. Here the
constraint $\operatorname{bin}\left[  2\underline{\underline{\Psi}%
}_{R,R^{\prime}},\frac{1}{2}\right]  +\operatorname{bin}\left[  2\underline
{\underline{\Psi}}_{R^{\prime},R},\frac{1}{2}\right]  =2\underline
{\underline{\Psi}}_{R,R^{\prime}}$\ applies in order to conserve agent number.
The second term of this attendance equation (Eq. \ref{Eq Attendance})
introduces a stochastic element in the game; it corresponds to the situation
where agents may have several top-scoring strategies and must thereby toss a
coin to decide which to use. We note that Eq. \ref{Eq Attendance} could be
re-written replacing the $\operatorname*{sgn}$ function with a $\tanh$. The
effect of this would be to make the number of agents playing strategy $R1$ (as
opposed to their other strategy $R2$) vary smoothly as a function of the
separation in score of the two strategies, rather than simply playing the
best. This modification is similar in concept to that of the Thermal Minority
Game (TMG) \cite{Sherrington_TMG,Us_TMG} wherein agents play their best
strategy with a certain probability depending on its score. The difference
here would be that, in contrast to the TMG, the system would still be entirely
deterministic hence lending itself readily to similar analysis as presented here.

With this formalism, the game can be described concisely by the following
coupled mapping equations:
\begin{align}
\underline{S}\left[  t\right]   &  =\underline{S}\left[  t-1\right]
-\underline{a}^{\mu\left[  t-1\right]  }\chi\left[  A\left[  \underline
{S}\left[  t-1\right]  ,\mu\left[  t-1\right]  \right]  \right]
\label{Eq Score Update}\\
\mu\left[  t\right]   &  =2\mu\left[  t-1\right]  -P\operatorname{H}\left[
\mu\left[  t-1\right]  -\frac{P}{2}\right]  +\operatorname{H}\left[  -A\left[
\underline{S}\left[  t-1\right]  ,\mu\left[  t-1\right]  \right]  \right]
\label{Eq Hist Update}%
\end{align}
where $\operatorname{H}\left[  x\right]  $ is the Heaviside function and
$\chi\left[  A\right]  $ is a monotonic, increasing function of the game
attendance quantifying the particular choice of reward structure (i.e.
payoff). \ In most of the MG literature $\chi\left[  A\right]
=\operatorname{sgn}\left[  A\right]  $ or $\chi\left[  A\right]  =A$
\cite{Savit_Original-MG}. Although the macroscopic statistical properties of
the MG are largely unaltered by the choice of $\chi$, we later demonstrate
that the microscopic dynamics can be affected markedly.

This formulation shows that the MG obeys a one-step, stochastically perturbed
deterministic mapping between states $\{\underline{S}\left[  t\right]
,\mu\left[  t\right]  \}$ and $\{\underline{S}\left[  t+1\right]  ,\mu\left[
t+1\right]  \}$. It is interesting to ask the following question: `How
important is the stochastic term of Eq. \ref{Eq Attendance} to the resultant
dynamics?'. Table 1 shows the frequency with which the outcome
($\operatorname{sgn}\left[  -A\right]  $) is changed by the stochastic
perturbation to the mapping. We can see that the stochastic term has a small
but non-negligible effect on the game. \ For the strategy reward system
$\chi=\operatorname{sgn}$, the number of instances of coin-tossing agents
affecting the outcome is greater than with the proportional reward system of
$\chi=1$. \ This is easily understood in terms of the homogeneity of the
score-vector $\underline{S}$; the $\chi=\operatorname{sgn}$ scoring system is
much more likely to generate tied strategies than the $\chi=1$ system which
also incorporates the dynamics of the attendance $A$. Therefore, in the
$\chi=\operatorname{sgn}$ scoring system there will be a much higher
proportion of coin-tossing agents and thus a greater effect on the game.

The general effect of the stochastic contribution to the MG is to break the
pattern of behavior emergent from the deterministic part of the map. \ It is
therefore of great interest to examine further what the dynamics of this
deterministic behavior are. To do this we replace the stochastic term of Eq.
\ref{Eq Attendance} by its mean. The equation thus becomes ($A_{D}\left[
\underline{S},\mu\right]  $ in Ref. \cite{Mike_TMG_Markov}):
\begin{equation}
A\left[  \underline{S},\mu\right]  =\sum_{R=1}^{2P}a_{R}^{\mu}\sum_{R^{\prime
}=1}^{2P}\left(  1+\operatorname{sgn}\left[  S_{R}-S_{R^{\prime}}\right]
\right)  \underline{\underline{\Psi}}_{R,R^{\prime}}\label{Eq Det Attendance}%
\end{equation}
Physically this replacement is an averaging process; when $S_{R1}=S_{R2}$\ we
have half the agents who hold $\{R1,R2\}$ playing $R1$ and the other half
playing $R2$ \cite{zero}. Equations \ref{Eq Score Update},\ref{Eq Hist Update}
\& \ref{Eq Det Attendance} now define a deterministic map which replicates the
behavior of the MG between perturbative events caused by the coin-tossing
agents - we refer to this system as the `Deterministic Minority Game' (DMG).
\ We will now use this system to investigate the emergence of microscopic and
macroscopic dynamics.

\bigskip

\section{Disorder in $\underline{\underline{\Psi}}\label{Sec Disorder}$}

The game is conditioned at the start with the initial state $\{\underline
{S}\left[  0\right]  ,\mu\left[  0\right]  \}$. It is also given a
$\underline{\underline{\Psi}}$ tensor for a particular parameter set $N,m,s$.
\ The game's future behavior will be inherited from $\underline{\underline
{\Psi}}$; games with sparsely and densely filled tensors hence behave in
entirely different ways. By assuming each entry of $\underline{\underline
{\Omega}}$ is an independent binomial sample $\underline{\underline{\Omega}%
}_{R1,R2}=\operatorname{bin}\left[  N,\frac{1}{\left(  2P\right)  ^{s}%
}\right]  $ we may categorize the disorder in the $\underline{\underline
{\Omega}}$\ tensor by the standard deviation of an element divided by its mean
size. For $s=2$, this gives
\[
\frac{\sigma\left[  \underline{\underline{\Omega}}_{R1,R2}\right]  }%
{\mu\left[  \underline{\underline{\Omega}}_{R1,R2}\right]  }=\sqrt
{\frac{\left(  2P\right)  ^{2}-1}{N}}%
\]
which rapidly becomes large as $m$ increases. \ For low $m$ and high $N$, the
game is said to be in an `efficient phase' \cite{EconPhys_Website} where all
states of the global information set $\mu$ are visited equally and hence, on
average, there is no drift in the strategies' scores i.e. $\left\langle
S_{R}\right\rangle _{t}=0$. In this regime, the disorder in the $\underline
{\underline{\Omega}}$\ tensor is small and thus all elements are approximately
of equal magnitude. This in turn implies that the dynamics of the game are
dominated by the movement of $\underline{S}$ rather than by the asymmetry of
$\underline{\underline{\Omega}}$. \ The attendance of the ($s=2 $) game here
reduces to
\begin{equation}
A\left[  \underline{S},\mu\right]  \thickapprox\frac{N}{4P^{2}}\sum_{R=1}%
^{2P}a_{R}^{\mu}\sum_{R^{\prime}=1}^{2P}\operatorname{sgn}\left[
S_{R}-S_{R^{\prime}}\right] \label{Eq Flat Attendance}%
\end{equation}
The second sum in Eq. \ref{Eq Flat Attendance} corresponds to a quantity
$q_{R}$ which is based on the rank of strategy $R$; specifically
$q_{R}=2P+1-2\rho_{R}$ where $\rho_{R}$ is the rank of strategy $R$, with
$\rho_{R}=1$ being the highest scoring and $\rho_{R}=2P$ being the lowest
scoring. Hence Eq. \ref{Eq Flat Attendance} becomes
\begin{equation}
A\left[  \underline{S},\mu\right]  \thickapprox\frac{N}{4P^{2}}\underline
{a}^{\mu}\cdot\underline{q}\ \ \ .\label{Eq Attendance Approx}%
\end{equation}
We now examine the increment in strategy score, \underline{$\delta S$}$\left[
t\right]  =\underline{S}\left[  t\right]  -\underline{S}\left[  t-1\right]  .$
For simplicity, we here assume the proportional scoring system of $\chi=1$.
Hence
\[
\underline{\delta S}=-\underline{a}^{\mu}A\left[  \underline{S},\mu\right]
\thickapprox-\frac{N}{4P^{2}}\underline{a}^{\mu}\left(  \underline{a}^{\mu
}\cdot\underline{q}\right)  \ \ .
\]
If we average over uniformly occurring states of $\mu$, we then have for each
strategy
\[
\left\langle \delta S_{R}\right\rangle _{\mu}\thickapprox-\frac{N}{4P^{2}}%
\sum_{R^{\prime}=1}^{2P}\left\langle a_{R}^{\mu}a_{R^{\prime}}^{\mu
}\right\rangle _{\mu}q_{R^{\prime}}%
\]
We now use the orthogonality of strategies in the RSS: $\frac{1}{P}\sum_{\mu
}a_{R1}^{\mu}a_{R2}^{\mu}=\{0$ for $R1\neq R2,1$ for $R1=R2,-1$ for
$R2=\overline{R1}\}$\ . This yields
\begin{equation}
\left\langle \delta S_{R}\right\rangle _{\mu}\thickapprox\frac{N}{2P^{2}%
}\left(  \rho_{R}-\rho_{\overline{R}}\right) \label{Eq Score Inc Rho}%
\end{equation}
where $\overline{R}=2P+1-R$ is the anti-correlated strategy to $R$. \ Equation
\ref{Eq Score Inc Rho} now shows us explicitly that strategies and their
anti-correlated partners attract each other in pairs. \ The magnitude of the
score increment is also of interest; for low $m$ and high $N$ the attractive
force is large, which will cause the strategies to overshoot each other and
thus perform a constant cycle of swapping positions. \ As we increase $m$ or
decrease $N$ the attractive force becomes weaker and so the score cycling
adopts a longer time-period; it eventually becomes too weak to overcome the
separate force arising from the asymmetry in $\underline{\underline{\Psi}}$.
Hence the system moves away from the strongly mean-reverting behavior in
$\underline{S}$.

We can investigate this change of regime further by examining $\left\langle
\delta S_{R}\right\rangle _{\mu}$ for finite disorder in $\underline
{\underline{\Psi}}$ \cite{efficient}. \ Again using the orthogonality of
strategies in the RSS, we have
\begin{align}
\left\langle \delta S_{R}\right\rangle _{\mu}  &  =\delta S_{R}^{bias}+\delta
S_{R}^{restoring}=\label{Eq Score Inc}\\
&  -\sum_{R^{\prime}=1}^{2P}\left(  \underline{\underline{\Psi}}_{R,R^{\prime
}}-\underline{\underline{\Psi}}_{\overline{R},R^{\prime}}\right) \\
&  -\sum_{R^{\prime}=1}^{2P}\left(  \operatorname{sgn}\left[  S_{R}%
-S_{R^{\prime}}\right]  \underline{\underline{\Psi}}_{R,R^{\prime}%
}+\operatorname{sgn}\left[  S_{R}+S_{R^{\prime}}\right]  \underline
{\underline{\Psi}}_{\overline{R},R^{\prime}}\right)  \ \ .\nonumber
\end{align}
Equation \ref{Eq Score Inc} has two distinct contributions. The first term
$\delta S_{R}^{bias}$ arises from disorder in $\underline{\underline{\Psi}}$
alone and is time-independent, representing a constant bias on the score
increment. \ The second term $\delta S_{R}^{restoring}$\ acts as a
mean-reverting force on the strategy score; its magnitude depends on how many
strategies lie between it and its anti-correlated partner (just as in Eq.
\ref{Eq Score Inc Rho}). Figure 1 illustrates this for a case where $S_{R}>0$;
here the net contribution to $\delta S_{R}^{restoring}$ is likely to be
negative as there are more contributing elements with a negative sign than
with a positive sign. The strategies $R^{\prime}\ni-\left|  S_{R}\right|
<S_{R^{\prime}}<\left|  S_{R}\right|  $ always contribute terms
$-\operatorname{sgn}\left[  S_{R}\right]  \left(  \Psi_{R,R^{\prime}}%
+\Psi_{\overline{R},R^{\prime}}\right)  $ to $\delta S_{R}^{restoring}$ and so
will always act as a mean-reverting component. Terms from strategies outside
this range will always be divided into equally sized positive and negative
groups as shown in Fig. 1. These groups will on average cancel out each
other's effect on the score increment.

We can model the average magnitude of each term in Eq. \ref{Eq Score Inc} by
using the same binomial representation for the elements of $\underline
{\underline{\Omega}}$ as before. The mean magnitude of the bias and restoring
force terms $\left\langle \left|  \delta S_{R}^{bias}\right|  \right\rangle
_{R}$ and $\left\langle \left\langle \left|  \delta S_{R}^{restoring}\right|
\right\rangle _{S_{R}}\right\rangle _{R}$ are thus approximately given as
follows:
\begin{align}
\left\langle \left|  \delta S_{R}^{bias}\right|  \right\rangle _{R}  &
\thickapprox\sqrt{\frac{N}{P\pi}\left(  1-\frac{1}{\left(  2P\right)  ^{2}%
}\right)  }\label{Eq Approx Inc Terms}\\
\left\langle \left\langle \left|  \delta S_{R}^{restoring}\right|
\right\rangle _{S_{R}}\right\rangle _{R}  &  \thickapprox\frac{N\gamma}%
{4P^{2}}\ \ .\nonumber
\end{align}
The term $\gamma$ enumerates the average net number of terms in $\delta
S_{R}^{restoring}$\ that act to mean revert $S_{R}$ i.e. the excess number of
terms with sign $-\operatorname{sgn}\left[  S_{R}\right]  $. Averaged over the
entire set of strategies, we have $\gamma=2P$. Figure 2 shows that our
approximate form for the average strategy score bias in Eq. \ref{Eq Approx Inc
Terms} is extremely good over the entire range of $\alpha=P/N$ whereas the
approximation of the restoring force term becomes progressively worse as
$\alpha$ is increased. This effect can be explained in terms of the `market
impact' of a strategy. The greater the number of agents using a strategy
$n_{R}$, the greater its contribution is to the attendance as can be seen from
Eq. \ref{Eq Attendance Basic}. As $n_{R}$ is increased above $n_{R^{\prime
}\neq R}$, the greater the probability becomes of the game outcome
($-\operatorname{sgn}\left[  A\right]  $) being opposed to $a_{R}^{\mu}$ and
hence strategy $R$ being penalized. This effect will arise if the quenched
disorder in $\underline{\underline{\Psi}}$ is such that more agents hold
strategy $R$ than $R^{\prime}\neq R$. As $\alpha$ is raised and the quenched
disorder in $\underline{\underline{\Psi}}$ grows, this effect will become
increasingly important. Hence it can be seen that $\underline{\underline{\Psi
}}_{R,R^{\prime}}$ and $\left\{  S_{R},S_{R^{\prime}}\right\}  $ are not
independent as assumed in obtaining Eq. \ref{Eq Approx Inc Terms}, but are
instead correlated through the effect of market impact; this correlation
becomes more significant as $\alpha$ is increased.

The nature of the correlation between $\underline{\underline{\Psi}%
}_{R,R^{\prime}}$ and $\left\{  S_{R},S_{R^{\prime}}\right\}  $ introduced by
market impact is non-trivial in form as can be seen from Fig. 3. We will not
discuss the details of an analytic reconstruction of $\underline
{\underline{\Psi}}_{\rho,\rho^{\prime}}$ here, but will instead simply note
some straightforward constraints on its form. Let us take the approximation
that on average the ranking of the strategies $\{\rho_{R}\}$ is given by the
ranking of their bias terms $\{\delta S_{R}^{bias}\}$. This will be true
\emph{on average} for a system described by Eq. \ref{Eq Score Inc}. We then
use the approximation that $\delta S_{R}^{bias}\backsim N\left[  0,\sqrt
{\frac{N}{2P}\left(  1-\frac{1}{4P^{2}}\right)  }\right]  $. Ordering the bias
terms resulting from samples drawn from this distribution, gives us that
$\underline{\underline{\Psi}}_{\rho,\rho^{\prime}}$\ satisfies
\[
\operatorname*{Erf}\left[  \frac{\left\langle \delta S_{\rho}^{bias}%
\right\rangle }{\sqrt{\frac{N}{P}\left(  1-\frac{1}{4P^{2}}\right)  }}\right]
=\frac{P-\rho}{P}%
\]
with $\delta S_{\rho}^{bias}$ given by $-\sum_{\rho^{\prime}=1}^{2P}\left(
\underline{\underline{\Psi}}_{\rho,\rho^{\prime}}-\underline{\underline{\Psi}%
}_{\overline{\rho},\rho^{\prime}}\right)  $ as in Eq. \ref{Eq Score Inc}. This
relation gives us an indication of how the rank of a strategy is affected by
its excess population, and is consistent with the form of $\underline
{\underline{\Psi}}_{\rho,\rho^{\prime}}$ as shown in Fig. 3. Note that in the
absence of market impact we would not be able to write down any equation
linking these parameters and Fig. 3 would be flat with no structure.

We have thus shown how market impact is profoundly manifest within the
structure of the MG \cite{Challet-Zhang_Original-MG}. In particular, Fig. 2
shows clearly that consideration of market impact is necessary in the
calculation of the transition point from efficient to inefficient regimes
\cite{Challet-Zhang_Original-MG}. The game enters the inefficient regime if
the magnitude of the bias term to the score increment (arising from disorder
in $\underline{\underline{\Psi}}$) exceeds the magnitude of the restoring
force term. We can calculate when \textit{on average} strategies begin to
drift by looking at when $\left\langle \left|  \delta S_{R}^{bias}\right|
\right\rangle _{R}=\left\langle \left\langle \left|  \delta S_{R}%
^{restoring}\right|  \right\rangle _{S_{R}}\right\rangle _{R}$ in Eq. \ref{Eq
Approx Inc Terms}. This occurs near $\alpha=\alpha_{c}\thickapprox\frac{\pi
}{4}$. This over-estimation of the transition point (which numerically occurs
in the DMG at around $\alpha_{c}=0.39$) could be corrected by taking into
account the non-flat structure of $\underline{\underline{\Psi}}_{\rho
,\rho^{\prime}}$. We would like to stress here that only \textit{on average}
does there exist a specific point at which the game passes from mean-reverting
to biased behavior (efficient to inefficient regime). Because the behavior of
the game is dictated by the disorder in \underline{\underline{$\Omega$}} and
not just by the specific parameters $N,m,s$ alone, a knowledge of $\alpha$ is
not enough information to classify the game as being in either the efficient
or inefficient regime. Therefore it seems arguable as to whether $\alpha_{c}$
is a `critical' value for any particular realization of this system.

Equation \ref{Eq Score Inc} can also yield insight into the dynamics in the
regime past the transition point. \ We were able to predict from Eq. \ref{Eq
Score Inc Rho} that in the efficient regime, pairs of anti-correlated
strategies would cycle around each other thus producing an ever changing score
rank vector $\underline{\rho}$. In the inefficient regime wherein the strategy
scores have appreciable bias, it would be natural to assume that
$\underline{\rho}$ would rapidly find a steady state as the strategy scores
diverged. \ This in fact does not happen; for example, consider the outermost
pair of strategies in the score-space (i.e. the current best, and its
anti-correlated partner the worst) at a point in the game. \ For these
strategies, Eq. \ref{Eq Score Inc} is given approximately by
\[
\left\langle \delta S_{R}\right\rangle _{\mu}\thickapprox-\sum_{R^{\prime}%
=1}^{2P}\left(  \underline{\underline{\Psi}}_{R,R^{\prime}}-\underline
{\underline{\Psi}}_{\overline{R},R^{\prime}}\right)  -\operatorname{sgn}%
\left[  S_{R}\right]  \sum_{R^{\prime}=1}^{2P}\left(  \underline
{\underline{\Psi}}_{R,R^{\prime}}+\underline{\underline{\Psi}}_{\overline
{R},R^{\prime}}\right) \ \ .
\]
Irrespective of the disorder in $\underline{\underline{\Psi}}$, we have
$\left|  \delta S^{bias}\right|  \lesssim\left|  \delta S^{restoring}\right|
$. It is thus likely that this strategy pair attract each other until at least
one other pair take their place as best/worst. This behavior will lead to a
non-stationary $\underline{\rho}$-vector, even in this regime.

The present analysis has described general properties of the game such as the
transition in behavior between efficient and inefficient regimes. It has also
shown that dynamical processes such as the changing nature of $\underline
{\rho}$ can be quantitatively explained purely in terms of the quenched
disorder in the strategy population tensor \underline{\underline{$\Omega$}}.

\section{Dynamics in $\mu$-space\label{Sec Mu Dynamics}}

The previous section was concerned with the behavior of the strategy score
vector $\underline{S}$, and often treated the dynamical variable $\mu$ as a
random process to be averaged over. This however glosses over the subtle and
very interesting dynamics of $\mu$ itself as dictated by Eq. \ref{Eq Hist
Update}. (References \cite{Challet_MG-Memory} and \cite{Mike_TMG_Markov} also
consider aspects of $\mu$ dynamics). To aid in our discussion, we note that
Eq. \ref{Eq Hist Update} describes a trajectory along the edges of a directed
de Bruijn graph $\operatorname{D}_{2}\left[  m\right]  $. Fig. 4 shows an
example of such a graph for $m=2$. As explained in the previous section, in
the efficient regime $\underline{S}$ is strongly mean reverting. This implies
that the set of states of the game $\left\{  \underline{S},\mu\right\}  $ is
finite. As the system is Markovian and deterministic, this in turn implies
that it must exhibit periodic behavior in this regime as return to a past
state would then be followed by the revisiting of the trajectory from that
state. In the inefficient regime where the strategy scores are biased, the set
of states $\left\{  \underline{S},\mu\right\}  $\ is unbounded and we may
expect aperiodic behavior of the DMG.

We now examine the structure of the periodic behavior in the efficient regime.
One observation from numerical simulations is that the period i.e. return time
to any state $\left\{  \underline{S},\mu\right\}  $ is observed over many runs
to be $T=2P$ for the $\chi=\operatorname*{sgn}$ scoring system whereas for the
$\chi=1$ system the period is much longer and run-dependent. This periodic
behavior seems able to exist up to the point where the occurrence of zero
attendance $A\left[  \underline{S},\mu\right]  =0$ causes stochastic
perturbation to $\mu$ \cite{zero}; after this point we can no longer treat our
system as deterministic. Such periodic behavior must satisfy the conditions
$\left\{  \underline{\Delta S}_{cycle}=0,\underline{\Delta\mu}_{cycle}%
=0\right\}  $. $T=2P$ is in fact the shortest possible period which satisfies
these conditions. The two edges leading away from any vertex $\mu$\ on the de
Bruijn graph must necessarily inccur score increments of opposite sign:
$+\underline{a}^{\mu}\left|  \chi\left[  A\right]  \right|  ,-\underline
{a}^{\mu}\left|  \chi\left[  A\right]  \right|  $\ corresponding to positive
and negative attendance respectively. The vectors \underline{$a$}$^{\mu1}$ and
\underline{$a$}$^{\mu2\neq\mu1}$ are orthogonal; hence the only way that an
increment to the score of $\underline{a}^{\mu\left[  t\right]  }\chi\left[
A\left[  \underline{S}\left[  t\right]  ,\mu\left[  t\right]  \right]
\right]  $ can be negated in order to achieve $\underline{\Delta S}_{cycle}%
=0$, is to return to that vertex (i.e. $\mu\left[  t^{\prime}\right]
=\mu\left[  t\right]  $) a particular number of times such that
\begin{equation}
\sum_{\left\{  t^{\prime}\right\}  }\chi\left[  A\left[  \underline{S}\left[
t^{\prime}\right]  ,\mu\left[  t\right]  \right]  \right]
=0\label{Eq Cycle Condition}%
\end{equation}
This condition must be satisfied at all vertices of the graph because the set
$\left\{  t^{\prime}\right\}  $ which satisfies Eq. \ref{Eq Cycle Condition}
must have a minimum of two entries (each of opposite attendance) thereby
leading the game to different, new vertices until all are spanned.

Consider the $\chi=\operatorname*{sgn}$ scoring system. The condition
corresponding to Eq. \ref{Eq Cycle Condition} is easily satisfied at each
vertex with a set $\left\{  t^{\prime}\right\}  $ of exactly $2\lambda$
entries, $\lambda$ being an integer. We now have the situation where all edges
of the graph are visited equally. The shortest way of doing this is with
$\lambda=1$; this cycle is known as an `Eulerian trail'. This dynamical stable
state of the game acts as an attractor; the MG in the efficient phase will
rapidly find this state after undergoing a stochastic perturbation. We note
that the Time-Horizon Minority Game \cite{Mike_TMG_Markov} exhibits similar
behavior for special values of the time horizon $\tau$. This trajectory of the
DMG along a Eulerian trail corresponds to the occurrence of perfect
anti-persistence in the $[A|\mu]$ time series. This anti-persistence has been
empirically observed in many studies of the MG
\cite{Challet-Zhang_Original-MG,Savit_Original-MG,Zheng_EfficientStats}.

Now consider the $\chi=1$ scoring system. The condition of Eq. \ref{Eq Cycle
Condition} is very much harder to achieve over all vertices as the dynamics of
$A$ are incorporated back into the score vector \underline{$S$} making the set
$\left\{  \underline{S},\mu\right\}  $ very much larger. This explains the
very much longer period of this game which, even over very long time windows,
can appear aperiodic. The Eulerian trail will still however be an attractor to
the dynamics within $\mu$-space, since the anti-persistence in $[A|\mu]$ is
still strong (in the efficient phase). It is not however perfect as was the
case for the DMG using the $\chi=\operatorname*{sgn}$ scoring system.

To quantitatively explain this anti-persistence, we make the following
approximation:
\begin{equation}
\operatorname*{sgn}\left[  A\right]  \thickapprox\operatorname*{sgn}\left[
\underline{a}\cdot\underline{S}\right]  \ \ .\label{Eq Sgn A Approx}%
\end{equation}
This approximation can be understood by referring back to Eq. \ref{Eq
Attendance Approx} where $\underline{S}$ now plays the same role as the
rank-measure $\underline{q}$. It is valid for the regime where the strategy
scores are densely spaced, i.e. for the efficient regime/low disorder in
$\underline{\underline{\Psi}}$. Consider the $\chi=\operatorname*{sgn}$
scoring system wherein the score vector is simply given by $\underline
{S}\left[  t\right]  =\underline{S}\left[  0\right]  -\sum_{j=1}%
^{t-1}\operatorname*{sgn}\left[  A\left[  j\right]  \right]  \underline
{a}^{\mu\left[  j\right]  }$. We use the fact that the vectors \underline{$a$%
}$^{\mu1}$ and \underline{$a$}$^{\mu2\neq\mu1}$ are orthogonal to transform
Eq. \ref{Eq Sgn A Approx} to the following form:
\begin{equation}
\operatorname*{sgn}\left[  A\left[  t\right]  \right]  \thickapprox
\operatorname*{sgn}\left[  \underline{a}^{\mu}\cdot\underline{S}\left[
0\right]  -2P\sum_{\left\{  t^{\prime}\right\}  }\operatorname*{sgn}\left[
A\left[  t^{\prime}\right]  \right]  \right] \label{Eq Antipersistence}%
\end{equation}
where we recall that the set of times $\left\{  t^{\prime}\right\}  $ are such
that $\mu\left[  t^{\prime}\right]  =\mu\left[  t\right]  =\mu$ for
$0<t^{\prime}<t$. This dynamical process occurring over times $t^{\prime}$
rather than $t$ is zero-reverting. Let us demonstrate this by taking an
example. Let $P=4$ and the initial strategy score be such that $\underline
{a}^{\mu}\cdot\underline{S}\left[  0\right]  =20$. The time-series of
$\operatorname*{sgn}\left[  A\left[  t\right]  \right]  $ thus becomes as
shown in Table 2. Hence the game cascades from its initial state, the
attendance at a given vertex of the de Bruijn graph ($[A|\mu]$) exhibiting
persistent behavior until a point is reached such that $\left|  \underline
{a}^{\mu}\cdot\underline{S}\left[  0\right]  -2P\sum_{\left\{  t^{\prime
}\right\}  }\operatorname*{sgn}\left[  A\left[  t^{\prime}\right]  \right]
\right|  <2P$. Subsequently the attendance $[A|\mu]$ becomes perfectly
\emph{anti}-persistent. When this anti-persistence occurs at each vertex, the
game has locked into one of the $2^{2^{m}}/2^{m+1}$ Eulerian trails. The
analysis above can be generalized for different scoring systems (such as
$\chi=1$) where in general it is found that the game exhibits strong but not
perfect anti-persistence in $[A|\mu]$ in this regime.

In the analysis above we introduced the effect of the initial condition on the
score vector $\underline{S}\left[  0\right]  $ (see also Ref.
\cite{Challet_StationaryStates}). However, we could just as correctly view
$\underline{S}\left[  0\right]  $ as the current state, left by some other
game process such as a shock to the system, a build up from some other game
mechanism or a stochastic perturbation. It is therefore interesting to examine
how the DMG evolves after a given state $\left\{  \underline{S}\left[
0\right]  ,\mu\left[  0\right]  \right\}  $ is imposed. The `initial'
condition $\underline{S}\left[  0\right]  $ must obey the form $S_{R}%
=-S_{\overline{R}}$; this is to ensure that a priori no strategies are given a
bias. It would be unphysical to break this rule; strategy $R$ always loses the
same number of points as its anti-correlated partner $\overline{R}$ gains in
any reasonable physical mechanism. We expect that if the elements
$S_{R}\left[  0\right]  $ have magnitude less than $2P$ then the system will
very quickly lock into the Eulerian trail trajectory and visit all $\mu
$-states equally. However, if the elements $\left|  S_{R}\left[  0\right]
\right|  \gg2P$ then Eq. \ref{Eq Antipersistence} predicts that there will be
persistence in $[A|\mu]$ until the dynamical stable state is found. This
persistence in trajectory at each node of the de Bruijn graph will lead to the
game visiting only a small subset of the vertices on the graph unlike in the
stable-state situation. This reduced cycling effect may lead to a bias in the
attendance over a significant period of time, i.e. a `crash' or `rally'.

We now demonstrate the recovery of the DMG from a randomly chosen initial
score vector $\underline{S}\left[  0\right]  $. We take a system with low
disorder in $\underline{\underline{\Psi}}$ and $m=2$ (such that $2P=8$).
However we draw $S_{R}\left[  0\right]  $ from a much wider uniform
distribution, spanning $-100...100$. (Note we maintain $S_{R}=-S_{\overline
{R}}$ as required). Figure 5 shows the evolution of the game out of this
state. The initial condition is soon `worked out' of the system - it rapidly
finds the Eulerian cycle $\mu=0,0,1,3,3,2,1,2,..$ after only 174 turns. As can
also be seen, the game adopts several different types of cycle on its way
towards this stable state. The switch between cycle types occurs as each
vertex snaps from persistent to anti-persistent behavior.

We have hence discussed and explained the dynamics of the stable state, and
how the system enters that state from an initial or perturbed state. This
analysis has been for the system in the efficient phase where the quenched
disorder of $\underline{\underline{\Psi}}$ is low. The inefficient regime will
in general show a different set of dynamics. As discussed earlier, the
inefficient phase is characterized by score vectors which have an appreciable
drift; this is an effect of the disorder in $\underline{\underline{\Psi}}$.
The corresponding unbounded $\underline{S}\left[  t\right]  $ vector leads to
an unbounded set of states for the system $\left\{  \underline{S},\mu\right\}
$. This suggests that the overall dynamics may be aperiodic, i.e. the system
never returns to a past state. We can however say something about the nature
of the resulting dynamics in $\mu$-space. As the score vector diverges the
score rank vector $\underline{\rho}$ becomes more well defined (although not
completely stationary in time, as mentioned in Sec. \ref{Sec Disorder}). This
is tantamount to there being a certain degree of persistence in the attendance
at a vertex $[A|\mu]$. This will lead to the motion around the de Bruijn graph
being limited to a certain sub-space, just as that described above for the
recovery from an initial score vector $\underline{S}\left[  0\right]  $. This
difference in the dynamics for the efficient and inefficient regimes leads to
the well-documented result that the occurrence of different $m+1$ bit words is
even in the efficient regime but very uneven in the inefficient regime
\cite{Savit_Original-MG}.

\section{Conclusion}

The results in this paper confirm that the MG can be usefully viewed as a
stochastically perturbed deterministic system, and that this deterministic
system can be described concisely by coupled mapping equations (Eqs. \ref{Eq
Score Update}, \ref{Eq Hist Update} and \ref{Eq Det Attendance}). We used this
system to explore the dynamics of the score vector $\underline{S}\left[
t\right]  $. We showed that the score increment comprises a bias and
restoring-force term, the comparative magnitude of these terms being governed
by the disorder in the strategy population tensor $\underline{\underline
{\Omega}}$. Furthermore, we were able to obtain analytic approximations for
the bias and restoring force terms. We showed how the market-impact effect
correlated the strategy population to the score vector and how this then
affected our approximations.

We also discussed the dynamics of the global information $\mu\left[  t\right]
$ as a trajectory on a de Bruijn graph. We were able to show that in the
efficient regime the system would be periodic and that the favored periodic
trajectory was that of an Eulerian Trail. Analytically we were able to
demonstrate how anti-persistence and persistence arise in the attendance at a
vertex $[A|\mu]$, and how this would manifest itself in efficient and
inefficient regimes either in response to a perturbed state or an initial
condition of $\underline{S}\left[  0\right]  $.

\vskip2cm

\bigskip

We are grateful to A. Short and P.M. Hui for useful discussions and comments.

\newpage

\bigskip\centerline{\bf TABLES}

\vskip1cm%

\begin{tabular}
[c]{|l|l|l|}\hline
m & $\chi=\operatorname{sgn}$ & $\chi=1$\\\hline
2 & $7.2\pm4.2$ & $0.7\pm0.6$\\\hline
3 & $6.3\pm3.0$ & $2.4\pm0.8$\\\hline
4 & $9.4\pm2.1$ & $3.4\pm0.8$\\\hline
\end{tabular}

\bigskip

TABLE 1. Percentage of time-steps in which the minority room is changed by the
stochastic decision of agents with tied strategies. Percentages are shown for
the digital and proportional payoffs. Statistics obtained from 16 numerical
runs of the MG with $N=101$, $s=2$, and over 1000 time-steps.

\vskip1cm%

\begin{tabular}
[c]{|l|l|}\hline
$\underline{a}^{\mu}\cdot\underline{S}\left[  0\right]  -2P\sum_{\left\{
t^{\prime}\right\}  }\operatorname*{sgn}\left[  A\left[  t^{\prime}\right]
\right]  $ & $\operatorname*{sgn}\left[  A\left[  t\right]  \right]  $\\\hline
$20-8\times\left(  0\right)  =20$ & $1$\\\hline
$20-8\times\left(  0+1\right)  =12$ & $1$\\\hline
$20-8\times\left(  0+1+1\right)  =4$ & $1$\\\hline
$20-8\times\left(  0+1+1+1\right)  =-4$ & $-1$\\\hline
$20-8\times\left(  0+1+1+1-1\right)  =4$ & $1$\\\hline
$20-8\times\left(  0+1+1+1-1+1\right)  =-4$ & $-1$\\\hline
\end{tabular}

\bigskip

TABLE 2. An example of how the game cascades from an initial state (c.f. Eq.
\ref{Eq Antipersistence}). Here $P=4$ and $\underline{a}^{\mu}\cdot
\underline{S}\left[  0\right]  =20$. The attendance (right column) exhibits
persistent, and then anti-persistent, behavior.

\newpage\centerline{\bf FIGURE CAPTIONS}

\vskip1cm

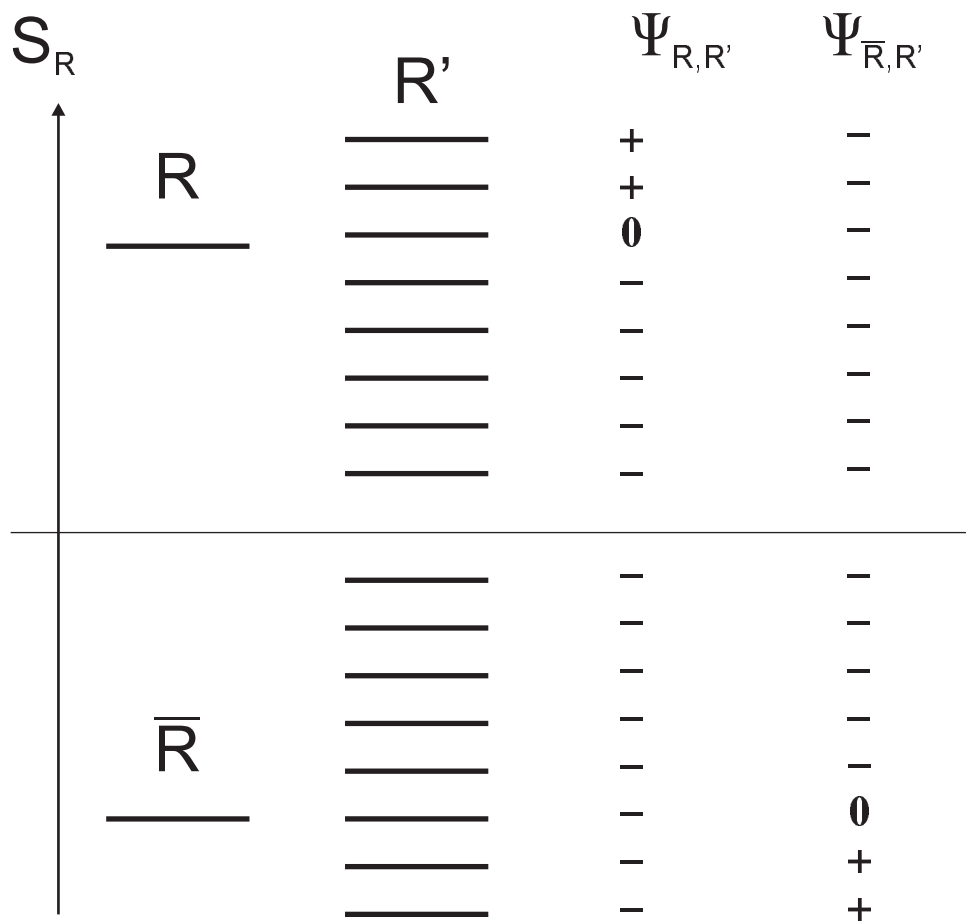
FIG. 1. Schematic representation of the signs of contributing terms to $\delta
S^{restoring}$.

\vskip1cm

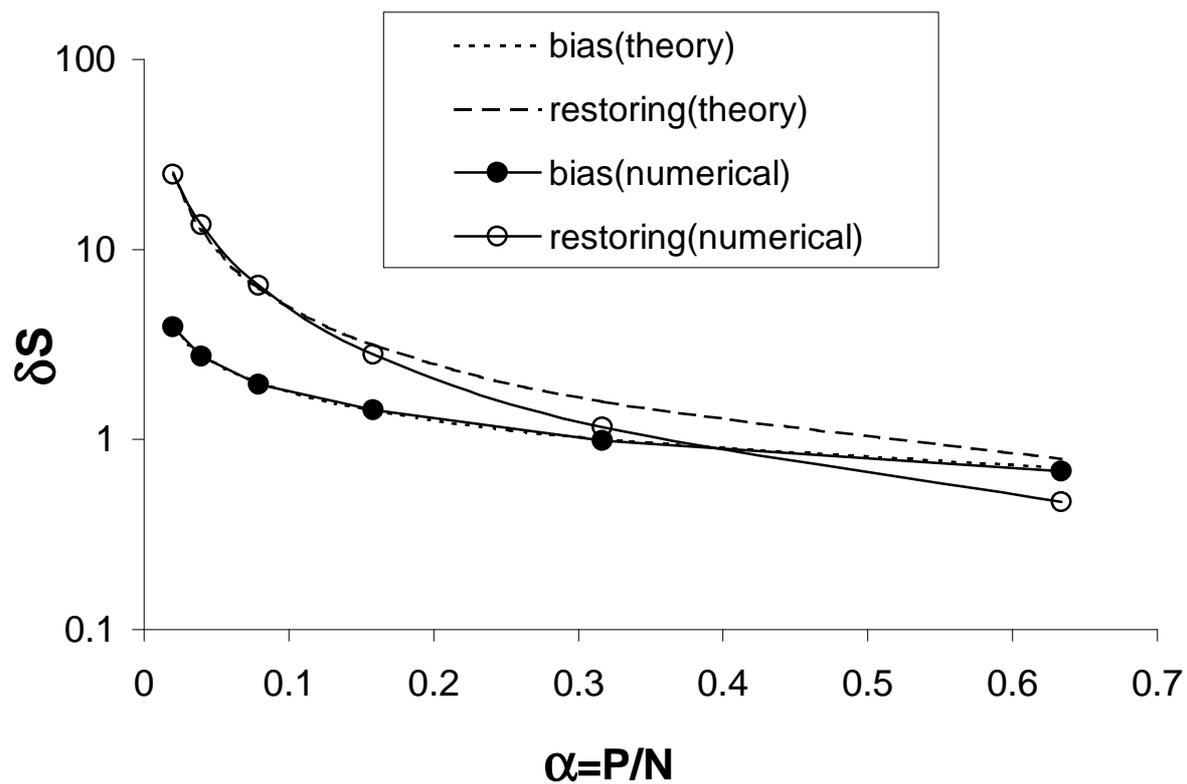
FIG. 2. Numerical and approximate analytical magnitude of average score
increment terms $\left\langle \left|  \delta S_{R}^{bias}\right|
\right\rangle _{R}$ and $\left\langle \left\langle \left|  \delta
S_{R}^{restoring}\right|  \right\rangle _{S_{R}}\right\rangle _{R}$

\vskip1cm

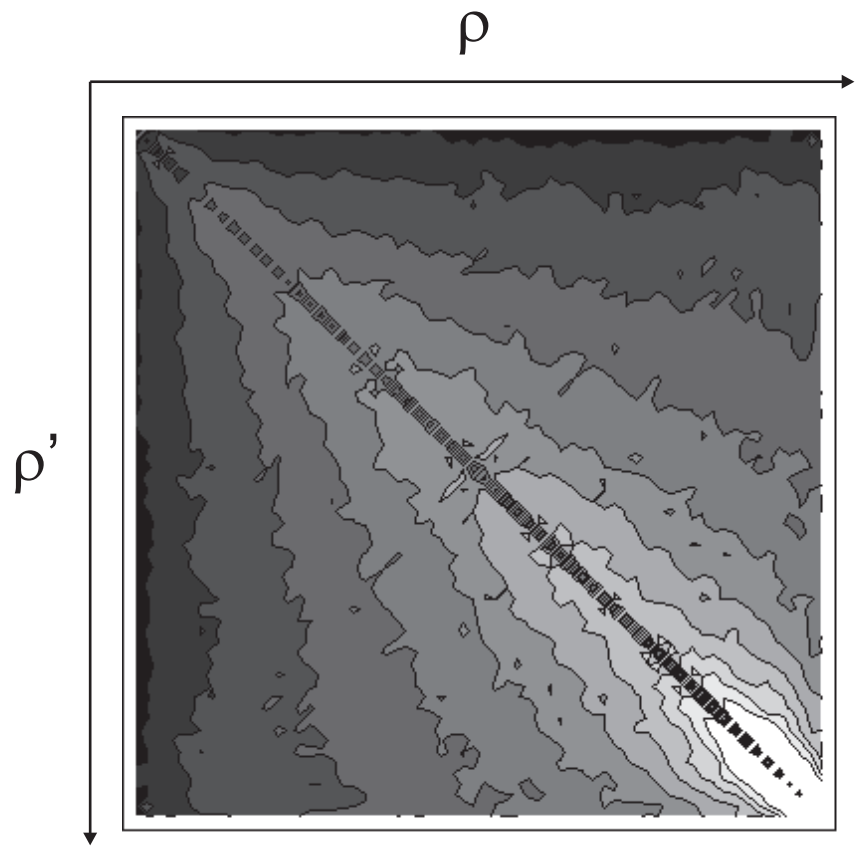
FIG. 3. Contour plot of $\left\langle \underline{\underline{\Psi}}_{\rho
,\rho^{\prime}}\right\rangle $, i.e. an average of the strategy population
tensor re-ordered each turn with strategies running from highest to lowest
score (top to bottom and left to right). Black areas indicate low population
and white areas indicate high population. The averaging is carried out over 50
runs (different $\underline{\underline{\Omega}}$) and 1000 turns within each
run. MG game parameters $\alpha=0.32$, $s=2$.

\vskip1cm

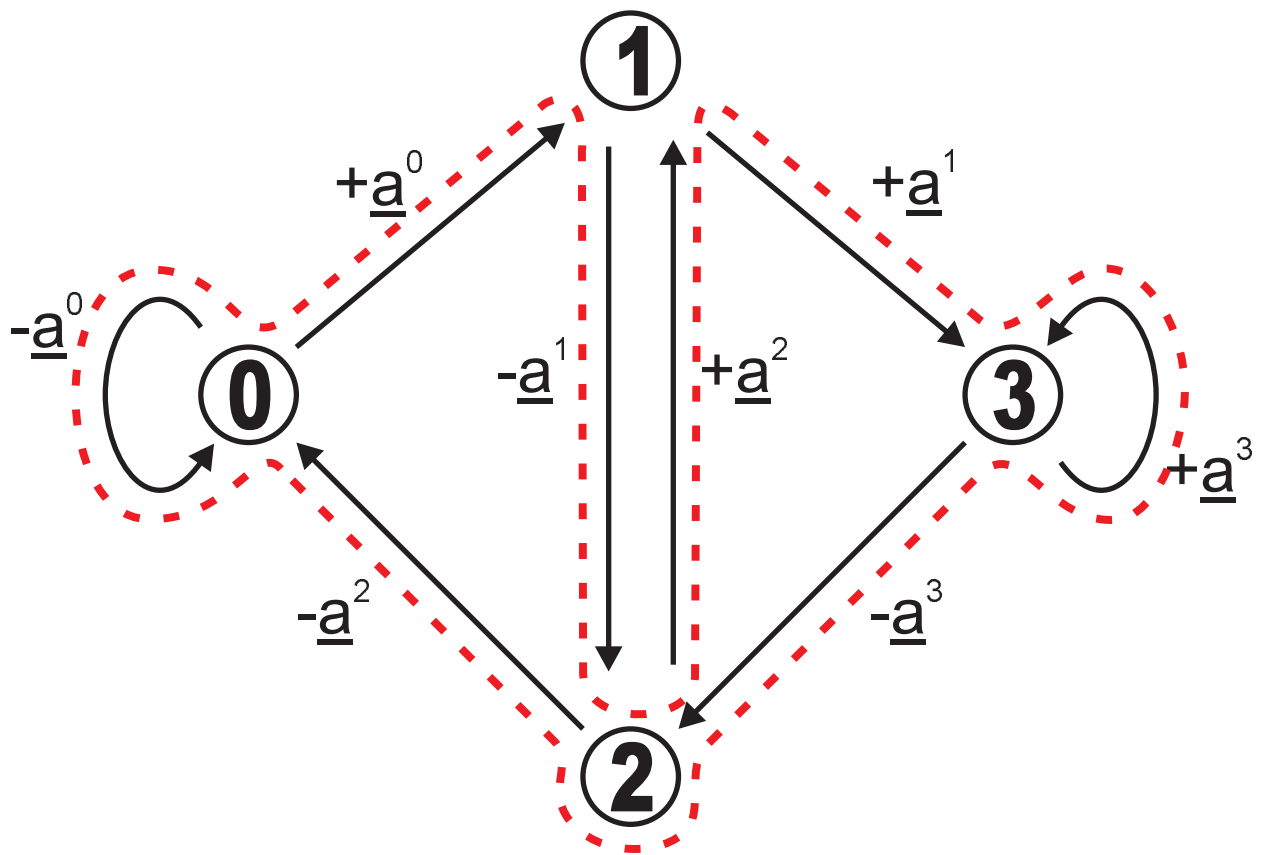
FIG. 4. De Bruijn graph $\operatorname{D}_{2}\left[  2\right]  $ corresponding
to $m=2$. Vertices are labelled with the state $\mu$, edges are labelled with
the quantity $\underline{\delta S}/\left|  \chi\left[  A\right]  \right|  $.
The dotted line shows one of the two possible Eulerian trails of this graph.

\vskip1cm

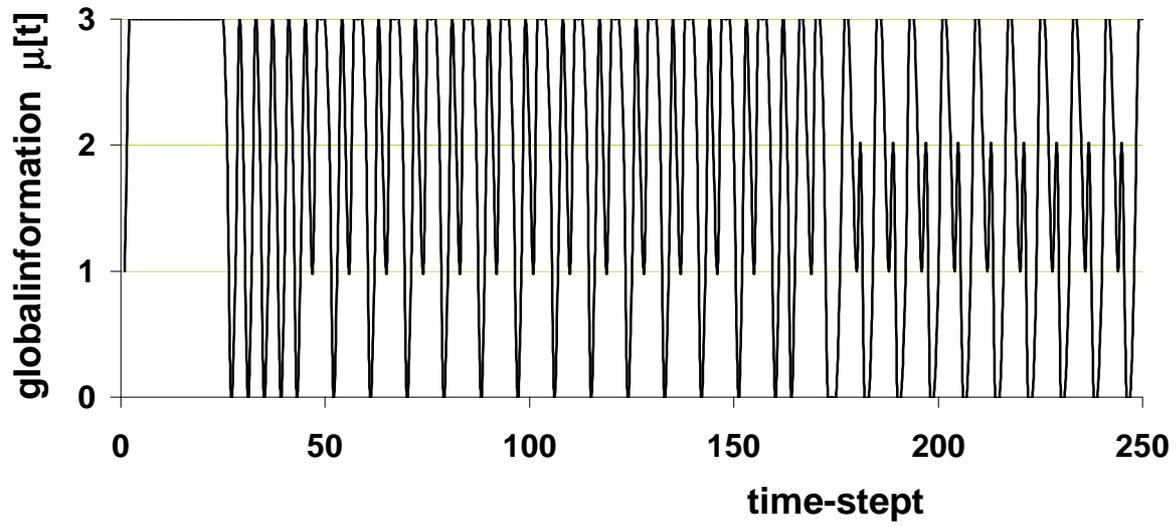
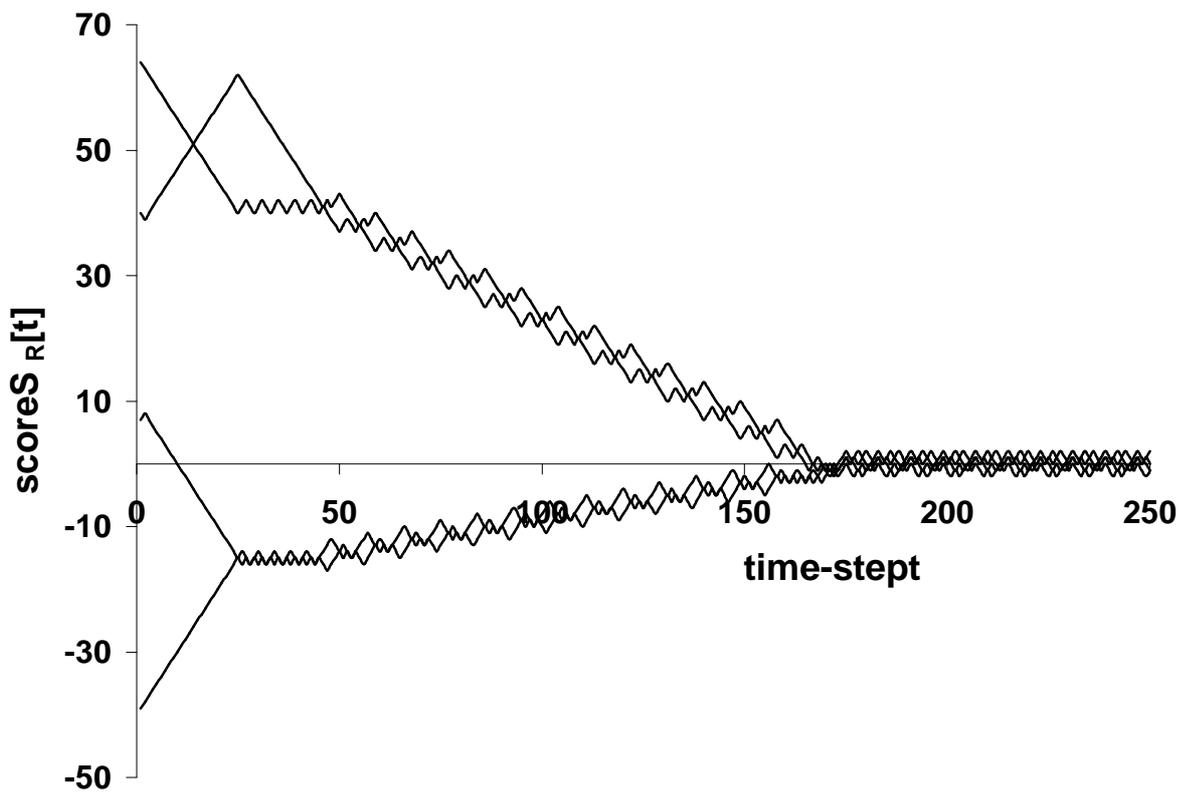
FIG. 5. An example of the convergence of the DMG onto the Eulerian trail
attractor. Top graph shows the dynamics in the global information $\mu\left[
t\right]  $. Bottom graph shows the dynamics in score $S_{R}\left[  t\right]
$ for $1\leqslant R\leqslant4$ (out of $2P=8$). Game locks into attractor at
turn 174. Game parameters $N=101$, $m=2$, $s=2$.

\begin{thebibliography}{99}
\bibitem{countpsi}The expression \underline{\underline{$\Omega$}}%
+\underline{\underline{$\Omega$}}$^{\text{T}}$ double-counts the diagonal
terms of \underline{\underline{$\Omega$}}. This property is compensated for by
the form of Eq. \ref{Eq Attendance}.
\bibitem{zero}Partitioning the number of agents with tied strategies in this
way, may lead to non-integer $n_{R}$. In turn a zero attendance is possible
even from an odd number of agents. In these (rare) occurrences we assign
$A=\pm1$ randomly. This break in the determinism of the system is of
particular relevance to the discussion in Sec. \ref{Sec Mu Dynamics}.
\bibitem{efficient}As the game is moved away from the efficient regime wherein
states of $\mu$ are visited equally, averaging over $\mu$ should strictly
become invalid. However some average properties of the game are reasonably
insensitive to the replacement of the $\mu$-process of Eq. \ref{Eq Hist
Update} by a uniform random process. See Ref.
\cite{Challet_MG-Memory,Cavagna_RandomHist} for a discussion of the validity
of random $\mu$.
\end{thebibliography}
\end{document}